\documentclass[twocolumn, prb,superscriptaddress]{revtex4}
\usepackage{amssymb}
\usepackage{graphicx}
\usepackage{dcolumn}
\usepackage{bm}
\usepackage{amsmath}

\begin{document}

\title{Exact ordering of energy levels for one-dimensional interacting Fermi
gases with $SU(n)$ symmetry}
\author{Lei Pan}
\thanks{These authors contribute the work equally}
\affiliation{Beijing National Laboratory for Condensed Matter Physics, Institute of
Physics, Chinese Academy of Sciences, Beijing 100190, China}
\affiliation{School of Physical Sciences, University of Chinese Academy of Sciences,
Beijing, 100049, China}
\author{Yanxia Liu}
\thanks{These authors contribute the work equally}
\affiliation{Beijing National Laboratory for Condensed Matter Physics, Institute of
Physics, Chinese Academy of Sciences, Beijing 100190, China}
\affiliation{Institute of Theoretical Physics, Shanxi University, Taiyuan 030006, P. R.
China}
\author{Haiping Hu}
\affiliation{Beijing National Laboratory for Condensed Matter Physics, Institute of
Physics, Chinese Academy of Sciences, Beijing 100190, China}
\affiliation{Department of Physics, The University of Texas at Dallas, Richardson, Texas
75080, USA}
\author{Yunbo Zhang}
\affiliation{Institute of Theoretical Physics, Shanxi University, Taiyuan 030006, P. R.
China}
\author{Shu Chen}
\thanks{corresponding author, schen@iphy.ac.cn}
\affiliation{Beijing National Laboratory for Condensed Matter Physics, Institute of
Physics, Chinese Academy of Sciences, Beijing 100190, China}
\affiliation{School of Physical Sciences, University of Chinese Academy of Sciences,
Beijing, 100049, China}
\affiliation{Collaborative Innovation Center of Quantum Matter, Beijing, China}

\begin{abstract}
Based on the exact solution of one-dimensional Fermi gas systems with $SU(n)$
symmetry in a hard wall, we demonstrate that we are able to sort the ordering of the lowest
energy eigenvalues of states with all allowed permutation symmetries, which
can be solely marked by certain quantum numbers in the Bethe ansatz
equations. Our results give examples beyond the scope of the generalized Lieb-Mattis theorem, which
can only compare the ordering of energy levels of states belonging to
different symmetry classes if they are comparable according to the pouring
principle. In the strongly interacting regime, we show that the ordering
of energy levels can be determined by an effective spin-exchange model and extend our
results to the non-uniform system trapped in the harmonic potential.
\end{abstract}

\maketitle

%\affiliation{Institute of Theoretical Physics, Shanxi University, Taiyuan}

\section{Introduction}
The experimental progress in trapping and
manipulating ultra-cold atomic systems provides an ideal platform for
studying novel phenomena which are not easily accessible in solid state
systems\cite{Bloch2008}. A typical example is that the ultra-cold fermion
system with large hyperfine spin can possess high symmetries of $SU(n)$ and
exhibit exotic quantum magnetic properties fundamentally different from the
large-spin solid state systems, which usually have only $SU(2)$ symmetry.
Large-spin alkaline and alkaline-earth fermion systems with $SU(n)$ symmetry
have already been experimentally realized in recent years \cite%
{Ottenstein2008,Huckans2009,Zhang2014,Scazza2014,Cappellini2014,Taie2010,Cazalilla2009,Christian,Pagano2014}%
.
%for example,the three-component $^{6}Li$ system with the emergence of a $SU(3)$ symmetry \cite{Ottenstein2008,Huckans2009}, the multicomponent quantum gas with emergence of $SU\left( n\right)$ symmetry was reported in $^{87}$Sr \cite{Zhang2014} and $^{173} $Yb atomic systems% \cite{Scazza2014,Cappellini2014,Taie2010}; the $SU\left( n\right) $ generalization of the Hubbard model was realized in optical lattices with alkaline-earth atoms\cite{Cazalilla2009,Taie2010,Christian}; the realization of repulsively interacting one-dimensional multicomponent ultracold fermions with $SU\left( n\right) $ symmetry was observed in \cite{Pagano2014}.
The experimental progress on ultracold Fermi gases has stimulated
considerable theoretical studies, which unveiled the high $SU(n)$ symmetry
can give rise to exotic properties in quantum magnetism and pairing
superfluidity \cite{Wu2005,Wu2014,Nataf,Chen2005}. Particularly, in the
strongly interacting limit, recent studies have shown that the
one-dimensional (1D) multicomponent gases can be effectively described by
spin-exchange models \cite%
{Deuretzbacher1,Deuretzbacher2,Deuretzbacher3,YangLi1,Amin,Lijun,Haiping,
YangLi2,Haiping2,Volosniev2,Volosniev3,Levinsen,Massignan,Volosniev2017,Deuretzbacher2014,Harshman2016}
and may be applied to study the quantum magnetism of spin system with $SU(n)$
symmetry \cite%
{Zhichao2014,Bois2015,Decamp2016,Nichael2016,Nataf2016B,Nataf2016B2,Yuzhu2016,Capponi}%
.

For an interacting spin-1/2 Fermi system with $SU(2)$ symmetry, the ground
state is the spin singlet state according to the well known Lieb-Mattis
theorem (LMT) \cite{Lieb}, which indicates $E(S)<E(S^{^{\prime }})$ if $%
S<S^{^{\prime }}$ for a system composed of $N$ electrons interacting by an
arbitrary symmetric potential, where $E(S)$ is the lowest energy of states
with total spin $S$. Such a theorem can not be applied to the system with $%
SU(n)$ symmetry as the energy levels are no longer able to be sorted by the
total spin $S$. Nevertheless, for the high-symmetry $SU(n)$ system, Lieb and
Mattis proved a theorem (hereafter referred as LMT II), that if $\alpha $
can be poured in $\beta $, then $E\left( \alpha \right) >E\left( \beta
\right) $, where $\alpha $ and $\beta $ are two different symmetry classes
and $E\left( \alpha \right) $ and $E\left( \beta \right) $ are the
respective ground-state energies of the two classes \cite{Lieb}.
%According to the pouring principle, there exist some symmetry classes for which one class can
%not be poured into another class and thus one is not able to compare their
%energy levels.
Obviously, there exist some symmetry classes beyond LMT II where one class
can not be poured into another and thus one is not able to compare their
energy levels directly. Recently, the LMT II for the trapped multicomponent
mixtures was tested by studying the 1D strongly interacting few-body Fermi
gases, showing that the ground state corresponds to the most symmetric
configuration allowed by the imbalance among the components \cite{Decamp}.

Although the LMT II generally can predict correctly the ground state for a
high-symmetry multicomponent system, it is still not clear whether the
ordering of energy levels can be solely sorted according to their symmetry
classes, especially for those symmetry classes which are not comparable by
the pouring principle \cite{Lieb}. Aiming to give some clues to the
question, in this work we study the repulsively interacting 1D
multicomponent Fermi gas with $SU(n)$ symmetry confined in a hard wall
potential. The model can be exactly solved using the powerful Bethe ansatz
(BA) method \cite{Sutherland1968,Gaudin}, permitting us to obtain the exact
energy spectrum for all allowed permutation symmetry classes corresponding
to various Young tableaux. Particularly, in the strongly interacting limit,
we show that the spin part is effectively described by an $SU(n)$
spin-exchange model and its coupling strength can be exactly derived from
the expansion of the ground state energy. Therefore the ordering of energy
levels can be determined by the effective spin-exchange model. We then
generalize our study to the system trapped in a harmonic trap, which can be
effectively described by an inhomogeneous spin-exchange model, and find that
the ordering of energy levels fulfils similar distributions as the exactly solvable
case.

\section{Model and results}
We consider the 1D $n$-component fermionic systems with $%
SU\left( n\right) $ symmetry which can be described by the following
Hamiltonian
\begin{equation}
H=\sum_{i}^{N}\left[ -\frac{\hbar ^{2}}{2m}\frac{\partial ^{2}}{\partial
x_{i}^{2}}+V\left( x_{i}\right) \right] +g\sum_{i<j}\delta \left(
x_{i}-x_{j}\right) .
\end{equation}%
Here $V\left( x_{i}\right) $ is the external potential, $g$ is the
zero-range two-body interaction strength. The interactions between different
components have the same coupling strength. For the spin-independent
interactions, the particle number of each spin component is conserved. First
we consider the exactly solvable case with $V\left( x_{i}\right) =0$ for $%
x_{i} \in (-L/2,L/2)$ and otherwise $V = \infty$ under the open boundary
condition $\Psi \left( x_{i}=\pm L/2\right) =0$. For convention, we
introduce the interaction strength $c=mg/\hbar ^{2}$ and use the natural
units $\hbar ^{2}=2m=1$ in the following calculation. The system can be
exactly solved by the BA method \cite{Sutherland1968,Gaudin,Guan2006} and
the corresponding Bethe ansatz equations (BAEs) are as follows:%
\begin{eqnarray}
2k_{j}L = 2\pi I_{j}-\sum_{\alpha =1}^{M_{1}}\left[ \theta \left( \frac{%
k_{j}-\lambda _{\alpha }^{1}}{c/2}\right) +\theta \left( \frac{k_{j}+\lambda
_{\alpha }^{1}}{c/2}\right) \right]  \label{log1}
\end{eqnarray}
with $j =1,2 \cdots ,N$ and
\begin{widetext}
\begin{eqnarray}
\sum_{\beta \neq \alpha }^{M_{r}}\left[ \theta \left( \frac{\lambda
_{\alpha }^{r}-\lambda _{\beta }^{r}}{c}%
\right) +\theta \left( \frac{\lambda _{\alpha }^{r}+\lambda
_{\beta }^{r}}{c}\right) \right]   \label{log2}
&=&2\pi J_{\alpha }^{r}+\sum_{\gamma =1}^{M_{r+1}}\left[
\theta \left( \frac{\lambda _{\alpha }^{r}-\lambda _{\gamma
}^{r+1}}{c/2}\right) +\theta \left( \frac{\lambda _{\alpha
}^{r}+\lambda _{\gamma }^{r+1}}{c/2}\right) %
\right] \notag  \\
& & +\sum_{\delta =1}^{M_{r-1}}\left[ \theta \left( \frac{\lambda
_{\alpha }^{r}-\lambda _{\delta }^{r-1}}{c/2}%
\right) +\theta \left( \frac{\lambda _{\alpha }^{r}+\lambda
_{\delta }^{r-1}}{c/2}\right) \right] ,
% \alpha  &=&1,\cdots ,M_{r},r=1,2,\cdots N-1,  \notag
\end{eqnarray}
\end{widetext}
with $\alpha=1,\cdots ,M_{r}$ and $r=1,2,\cdots n-1$, where $\lambda
_{\delta }^{0}=k_{\delta }$, $\theta \left( x\right) =2\arctan x$, $M_{0}=N$%
, $M_{n}=0$ and $M_{r}$ takes integer in descending order $M_{0} > M_{1} >
\cdots >M_{r}$. The quantum numbers $I_{j}$ and $J_{\alpha }^{r}$ are
integers. \textbf{\emph{${k_{j}}s$}} are quasi-momentum and $\lambda
_{\alpha }^{r}$ denote the spin rapidities which are introduced to describe
the motion of spin waves. The particle number $n_{r}$ in each spin component
connects with $M_{r}$ via the relation $n_{r}=M_{r-1}-M_{r}$, where we have
assumed the components are ordered so that $n_1 \geq n_2 \geq \cdots \geq
n_n $. For the repulsive case with $c>0$, there is no charged bound state
and the quasi-momenta $\{k_{j}\}$ take real values. The eigenvalue is given
by $E = \Sigma _{j=1}^{N}k_{j}^{2}$.
\begin{figure}[tbp]
\includegraphics[width=0.5\textwidth]{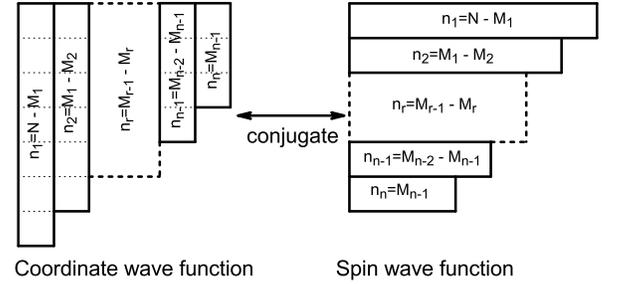}
\caption{Young tableau for the state of the $SU(n)$ $N$ -particle system
with given quantum numbers $M_{r}$. The left and right Young tableaus
represent symmetry classes of the coordinate and spin wave functions,
respectively, which are conjugated with each other.}
\label{fig1}
\end{figure}

\begin{figure}[tbp]
\caption{(color online). Ground state energies of states with different
permutation symmetries versus the dimensionless interaction strength $%
\protect\gamma $ for the $SU(4)$ system with particle number (a) $N=4$ and
(c) $N=6$. (a) From dash-dot-dot line to solid line corresponding to the
lowest-energy states with different symmetry classes described by the Young
tableaux of the coordinate wave function in (b). (c) From the solid line to
the dash-dot line corresponding to different symmetry classes described by
the Young tableaux shown in (d). }
\label{fig2}\includegraphics[width=0.48\textwidth]{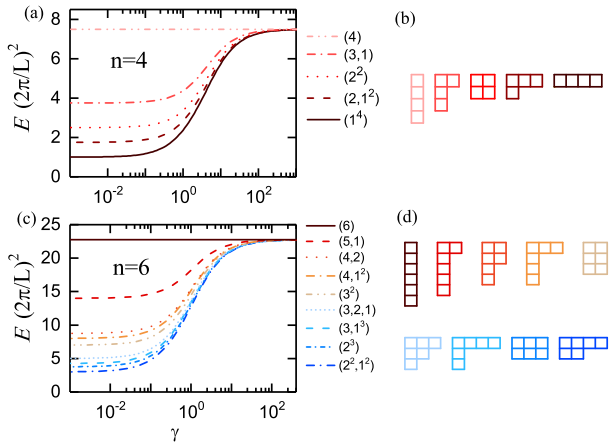}
%The particle numbers $n_{r}$ in the each spin component are equal to numbers in the box $\square $ of the Young diagram of the coordinate wave function in every column for the lowest-energy state.
\end{figure}

For a given set of $\{M_{r}\}$ ($r=0,\cdots ,n-1$), there exists a unique
Young tableau corresponding to the unique set of particle number
distribution $\{n_r\}$ (See Fig.1). Taking the case of $N=4$ with $SU(4)$
symmetry as an example, $\{M_{r}\}$ ($r=0,\cdots ,3$) can have five
different configurations, i.e, $\{M_{r}\}=\{4,3,2,1\}$, $\{4,2,1,0\}$, $%
\{4,2,0,0\}$, $\{4,1,0,0\}$ and $\{4,0,0,0\}$, correspondingly there exist
five sets of $\{n_{r}\}$, say, $\{n_{r}\}=\{1,1,1,1\}$, $\{2,1,1\}$, $%
\{2,2\} $, $\{3,1\}$ and $\{4\}$, which belong to different symmetry classes
with their eigenfunctions described by the Young tableaus in Fig.\ref{fig2}%
b, denoted by the simplified notations: $Y=(1,1,1,1)\equiv (1^{4})$, $%
(2,1,1)\equiv (2,1^{2})$, $(2,2)\equiv (2^{2})$, $(3,1)$ and $(4)$, from
right to left, respectively.
%Here, for example, the distribution $\{M_r\}=\{4,3,2,1\}$ gives rise to the occupation numbers $n_{1}=n_{2}=n_{3}=n_{4}=1$ ($\{n_{r}\}=(1,1,1,1)$), which corresponds to the Young tableau $\square\square\square\square$.
Here $n_{r}$ $(r=1,\cdots ,4)$ in $(n_{1},n_{2},n_{3},n_{4})$ indicates the
number of squares in the $r$-th column of the Young tableau, and the square
numbers in the $r$-th column are equal to the particle numbers $n_{r}$ of
the $r$-th component. For example, $(1,1,1,1)$ describes the Young tableau $%
\square \square \square \square $, which corresponds to the system with the
component-dependent particle numbers $n_{1}=n_{2}=n_{3}=n_{4}=1$.
% and the Young tableau of the spin wavefunction is conjugate with the space wave function's.
We note that the notation adopted here is different from the standard
notation of representation of Young tableau, but the same with Ref. \cite%
{Yang2009}.

According to the LMT II for the high-symmetry system, one can compare the
ground state energies of different symmetry classes if they fulfill the
pouring principle.
%Before discussing the ordering of energy levels, we would like to explain the meaning of pouring principle.
When we say that $\alpha $ can be poured in $\beta $, where the Young
tableau $\alpha $ has the columns $n_{1}\geq n_{2}\geq n_{3}\geq \cdots $
and the $\beta $ has the columns $n_{1}^{\prime }\geq n_{2}^{\prime }\geq
n_{3}^{\prime }\geq \cdots $, it means that we have $n_{1}\geq n_{1}^{\prime
}$; $\left( n_{1}-n_{1}^{\prime }\right) +n_{2}\geq n_{2}^{\prime }$; $%
\left( n_{1}-n_{1}^{\prime }\right) +\left( n_{2}-n_{2}^{\prime }\right)
+n_{3}\geq n_{3}^{\prime }$; $et.$ $al.$, here any missing columns are to be
regarded as having $n=0$ \cite{Lieb}.
%For the two-component case, total spin $S$ can \textbf{\emph{totally characterize}} the energy levels, since the Young tableau and total spin $S$ share the same eigenstates, i.e. the lowest or ground-state energy belonging to the given value $S(S^{^{\prime }})$satisfy $E\left( S\right) <E(S^{^{\prime }})$ if $S<S^{^{\prime }}$.
Different from the $SU(2)$ system, for the $SU(n)$ system, some symmetry
classes are not comparable by the pouring principle, e.g., symmetry classes denoted by $(3,3)$ and $(4,1^2)$ for the $SU(4)$
system with $N=6$.

Due to the unique correspondence between the symmetry classes and quantum
numbers in the BAEs, we can calculate the lowest eigenenergy for each given
symmetry class by numerically solving the corresponding BAEs with the
quantum numbers $\left\{ I_{j}\right\} =\left\{ 1,2,\cdots , N\right\} $ and
$\{J_{\alpha }^{r}\}=\left\{ 1,2,\cdots ,M_{r}\right\} $, which permits us
to determine the order of energy levels with different symmetries. We
present our results for the spin-$3/2$ fermionic gas with $SU\left( 4\right)
$ symmetry in Fig.\ref{fig2}. For the case of $N=4$ with five sets of $%
\{M_{r}\}$, the corresponding lowest energies for each symmetry classes are
shown in Fig.\ref{fig2}(a), indicating that the order of the energy levels
fulfills $E\left( 4\right) > E\left( 3,1\right) >E\left( 2,2\right)>E\left(
2,1^2\right) > E\left( 1^4\right)$ in the whole regime $0\leq \gamma <\infty
$, except in the Tonks-Girardeau limit $\gamma \rightarrow \infty $ all the
levels approach the same value, where $\gamma=cL/N$ is the dimensionless
interaction strength. We note that in this case, the order of energy levels
can also be determined by applying the LMT II, as all the five different
symmetry classes are comparable according to the pouring principle. However,
for the case of $N=6$, there exist incomparable symmetry classes, as
discussed before, neither $\left( 4,1^{2}\right) $ nor $\left( 3^{2}\right) $
can be poured into each other, so the LMT cannot determine the order of $%
E\left( 4,1^{2}\right) $ and $E\left( 3^{2}\right) $. Also, $E\left(
3,1^{3}\right) $ and $E\left( 2^{3}\right) $ are not comparable by the
pouring principle. Similar to the case of $N=4$, all the symmetry classes
for $N=6$ can be uniquely determined by solving the BAEs and thus we can
give an order of the energy levels as demonstrated in Fig.\ref{fig2}(c). 
%It is quite clear that all the energy levels can be ordered by their symmetry classes and
From our calculated results, we can determine the order  of energy levels of the incomparable symmetry classes 
and we have $E\left( 4,1^{2}\right) >E\left( 3^{2}\right) $ and $%
E\left( 3,1^{3}\right) >E\left( 2^{3}\right) $.
%Such a result obviously goes beyond the LMT II.
As shown in Fig.2 (c), the order of ground state energies with different
symmetries does not change for arbitrary finite interaction strength and no
phase transition occurs in the whole repulsive interaction region. 

%Therefore it is convenient and instructive to see the order of energy levels from the noninteracting limit of $c\rightarrow 0$ by counting the occupation of single-particle orbits. Due to the Pauli exclusion principle, two fermionswith same spin component occupying an identical orbit is forbidden. For a
%given distribution $\{n_{r}\}$, the energy is given by $E =
%\sum_{i=1}^{r}\sum_{l=1}^{n_{i}}\left( \frac{l\pi }{L}\right) ^{2}$, from
%which we can easyly find $E\left( 4,1^{2}\right) =32\pi ^{2}/L^{2}$ and $%
%E\left( 3^{2}\right) =28\pi ^{2}/L^{2}$, giving rising to $E\left(
%4,1^{2}\right) >E\left( 3^{2}\right) $.
\begin{figure}[tbp]
\caption{(color online). (a) The lowest energy versus $1/\protect\gamma $
for various symmetry classes. The scattered symbols represent results
obtained from the effective spin chain, whereas the solid lines are obtained
by solving the BAES. (b) The lowest energy calculated by exact
diagonalization of effective spin chain model of the $SU(4)$ system with $N=6
$ in a harmonic trap for various symmetry classes. Here we use the Young
tableaus of spin wave function, which conjugate with the tableaus of
coordinate wave function in Fig.2(d), to represent different symmetry
classes. }
\label{fig3}\includegraphics[width=0.45\textwidth]{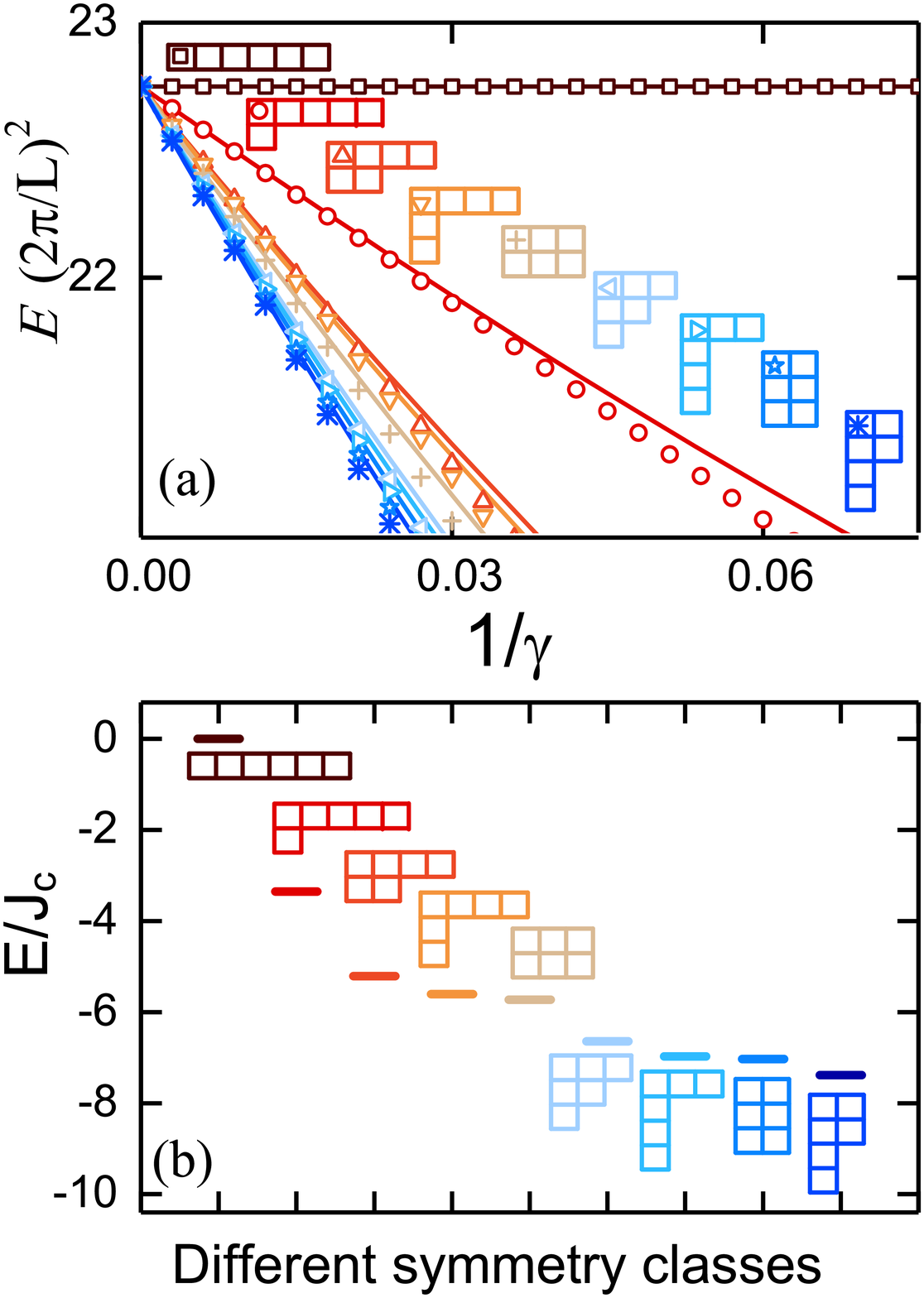}
%The particle numbers $n_{r}$ in the each spin component
%are equal to numbers in the box $\square $ of the Young diagram of the spin
%wave function in every row for the lowest-energy state.}
\end{figure}

%\section{The strongly interacting one-dimensional atomic gases}
%\textit{Strongly interacting regime.---}
In the strongly repulsive regime $%
\left( cL/N\gg 1\right) $, spin rapidities $\lambda _{\alpha }^{\left(
r\right) }$ are proportional to $c$ while $k_{j}$ remains finite. From the
expansion of Eq.(\ref{log1}) up to the first order in $k_{j}/c$, the
quasi-momentum is given by $2k_{j}L=2\pi I_{j}-2\zeta \frac{k_{j}}{c}%
+O\left( \left\vert c\right\vert ^{-3}\right)$, %\begin{equation}
%2k_{j}L=2\pi I_{j}-2\zeta \frac{k_{j}}{c}+O\left( \left\vert c\right\vert
%^{-3}\right),
%\end{equation}%
which leads to $k_{j}=\frac{\pi }{L}I_{j}\left( 1+\frac{1}{cL}\zeta \right)$
with
\begin{equation}
\zeta =\sum_{\alpha =1}^{M_{1}}\frac{4}{1+4 (\gamma_{\alpha }^{1})^2} .
\end{equation}
Here $\gamma_{\alpha }^{r} \equiv \lambda _{\alpha }^{r}/c$ and $%
\gamma_{\alpha }^{1}$ is determined by the following equation
\begin{eqnarray}
2N\theta \left( 2\gamma _{\alpha }^{1}\right) &=& 2\pi J_{\alpha
}^{1}-\sum_{\beta \neq \alpha }^{M_{1}} \left[ \theta \left( \gamma _{\alpha
}^{1}-\gamma _{\beta }^{1}\right) +\theta \left( \gamma _{\alpha
}^{1}+\gamma _{\beta }^{1}\right) \right]  \notag \\
& & -\sum_{\gamma =1}^{M_{2}}\left[ \theta \left( 2\gamma _{\alpha
}^{1}-2\gamma _{\gamma }^{2}\right) + \theta \left( 2 \gamma _{\alpha
}^{1}+2\gamma _{\gamma }^{2} \right) \right],  \label{spin1}
\end{eqnarray}%
which is obtained from the expansion of the second BAEs, i.e., Eq.(\ref{log2}%
) with $r=1$, up to the first order in $k_{j}/c$.
%which is the eigen-energy of the Heisenberg chain \cite{Sutherland1997}.
%Here we have used%
%\begin{equation*}
%\frac{d\tan ^{-1}\left( x\right) }{dx}=\frac{1}{1+x^{2}}\text{;}
%\end{equation*}%
%\begin{equation*}
%\theta \left( \frac{2\lambda _{\alpha }^{1}}{c}\right) +\theta \left( \frac{%
%-2\lambda _{\alpha }^{1}}{c}\right) =0.
%\end{equation*}%
The above equation can not solely determine $\gamma_{\alpha }^{1}$ as it
includes $\gamma_{\alpha }^{2}$, which should be iteratively determined by
the following equations
\begin{eqnarray}
&&\sum_{\beta \neq \alpha }^{M_{r}}\left[ \theta \left( \gamma _{\alpha
}^{r}-\gamma _{\beta }^{r}\right) +\theta \left( \gamma _{\alpha
}^{r}+\gamma _{\beta }^{ r}\right) \right] = 2\pi J_{\alpha }^{r} +  \notag
\\
&& ~ \sum_{\gamma =1}^{M_{r+1}}\left[ \theta \left( 2 \gamma _{\alpha
}^{r}-2 \gamma _{\gamma }^{r+1}\right) +\theta \left( 2 \gamma _{\alpha
}^{r}+2 \gamma _{\gamma }^{ r+1}\right) \right] +  \notag \\
& & ~ \sum_{\delta =1}^{M_{r-1}}\left[ \theta \left( 2 \gamma _{\alpha
}^{r}-2 \gamma _{\delta }^{r-1}\right) + \theta \left( 2 \gamma _{\alpha
}^{r}+ 2 \gamma _{\delta }^{r-1}\right) \right] ,  \label{spin2}
\end{eqnarray}%
with $\alpha =1,\cdots ,M_{r}$ ($r=2,\cdots n-1$).
% where $\gamma _{\alpha}^{r}=\lambda _{\alpha }^{r}/c$.
%It follows that the ground state energy of SU(n) Fermi gas in the strongly repulsive limit is given by
%\begin{eqnarray}
%E = \frac{\hbar ^{2}\pi ^{2}}{2mL^{2}}\frac{2N^{3}+3N^{2}+N}{6}\left( 1+%
%\frac{\zeta }{Lc}\right) ^{-2}+O\left( \left\vert c\right\vert ^{-3}\right) .
%\end{eqnarray}%
%
%
%Here we have used%
%\begin{equation*}
%\sum_{j=1}^{N}I_{j}^{2}=\frac{2N^{3}+3N^{2}+N}{6}.
%\end{equation*}%
%The quantum numbers $I_{j}$ of ground state are $\left\{ 1,2,\cdotsN-1,N\right\} $.
Up to the order of $c^{-1}$, the ground state energy of the $SU(n)$ Fermi
gas is given by
\begin{equation}
E=E_{F} \left(1 -\frac{2}{c L} \zeta \right) = E_{F} \left(1 - 2 \frac{%
\zeta/N }{\gamma} \right) ,  \label{energy}
\end{equation}%
where $E_{F}=\frac{\hbar ^{2}\pi ^{2}}{2mL^{2}}\frac{2N^{3}+3N^{2}+N}{6} $
is the energy at $c=\infty$. It is equal to the Fermi energy of fully
polarized Fermi gas, consistent with that from the generalized Bose-Fermi
mapping \cite{Guan2006,Yang2009,Chen2009,Deuretzbacher,Girardeau1,Girardeau2007}.

We note that Eq.(\ref{spin1}) and (\ref{spin2}) are the well-known Bethe
equations for the open $SU(n)$ Heisenberg spin chain
\begin{equation}
H_{S}=J\sum_{i=1}^{N-1}(P_{i,i+1}-1),
\end{equation}%
here $P_{i,i+1}$ is the permutation operator which permutes the spin states
of the $i$-th and $i+1$-th particles. The ground state energy of $H_{S}$
with $J>0$ is given by $E_{S}=-J\zeta $ \cite{Sutherland1975}.
%So according to the expression of Eq. \ref{energy} in the strongly repulsive
%regime,
By comparing with Eq.(\ref{energy}), we see that the effective Hamiltonian
describing the spin dynamics in the strongly interacting limit is given by $%
H_{eff}=H-E_{F}=H_{S}$ with the exchange parameter given by $J=2\frac{%
\epsilon _{F}}{\gamma }$, where $\epsilon _{F}=E_{F}/N$ is the average Fermi
energy. In Fig.\ref{fig3}(a), we show the energy spectrum of $H$ in the
strongly repulsive limit for the uniform system with $N=6$ by using the
effective Hamiltonian $H=E_{F}+H_{S}$. In the infinitely repulsive limit $%
\gamma \rightarrow \infty $, all the energy levels of different symmetry
classes are degenerate. Since the ground state energies of $H_{S}$ for
systems belonging to different symmetry classes take different values, the
energy levels split when the interaction strength deviates the TG limit. The
order of the splitting levels are convenient to be distinguished by the
Young diagrams of the spin wavefuntion, which are the conjugations of the
Young diagrams of the coordinate wavefuntion (see Fig.1).
%The effective spin-chain model can reproduce the exact low-energy spectrum of the exact Hamiltonian.

In the strongly interacting regime, the $SU(n)$ Fermi gas in an
inhomogeneous potential can also be described by a non-uniform effective
spin-chain model %\cite{YangLi1,Guan}
\begin{equation}
H_{eff}=\sum_{i=1}^{N-1}J_{i}(P_{i,i+1}-1),
\end{equation}%
with the coefficients given by 
\[
J_{i}=\frac{2N!}{c}\int
\prod_{j}dx_{j}|\partial _{i}\varphi _{A}|^{2}\delta (x_{i}-x_{i+1})\theta
_{\lbrack i,i+1]}^{1} ,
\] 
where $\theta _{\lbrack i,i+1]}^{1}=\theta
^{1}/\theta (x_{i}-x_{i+1})$ is a reduced sector function, $\theta ^{1}$
being the Heaviside step function whose value is $1$ in the region $%
x_{1}<x_{2}<\cdots <x_{N}$ and zero otherwise \cite{YangLi1,Lijun}. The wave
function $\varphi _{A}$ is taken as the ground state of $N$ spinless
fermions, i.e., the Slater determinant made up of the lowest $N$-level of
eigenstates. The difference from the uniform system is that the exchange
coefficients are site-dependent. A generalization of LMT for the $SU(n)$ chain is given
in Ref.\cite{Hakobyan}. Consider the $SU(4)$ system with $N=6$ in a
harmonic trap $V(x)=m\omega ^{2}x^{2}/2$ with the trapping frequency $\omega
$, we get a nonuniform $SU(4)$ spin chain with $J_{3}\equiv J_{c}$, $%
J_{1}=J_{5}=0.5743J_{c}$ and $J_{2}=J_{4}=0.8956J_{c}$, where $J_{c}$
represents the effective exchange strength between two spins in the trap
center. By directly diagonalizing the corresponding spin chain model, we can
get the order of energy levels for the harmonic system in the strong
interaction strength region. As shown in Fig. \ref{fig3}(b), the order of
energy levels is solely related to their symmetry classes, which agrees with
the case of uniform system. Our result indicates that the order of energy levels in the strongly interacting regime is not changed when the trap potential is changed from the hard wall to the harmonic trap.
Strongly interacting systems trapped in other external traps can be also studied similarly by solving corresponding effective spin-exchange models.
%Our result indicates that the change of trap potential does not change the order of energy levels of different symmetry classes.

\begin{figure}[tbp]
\includegraphics[width=0.45\textwidth]{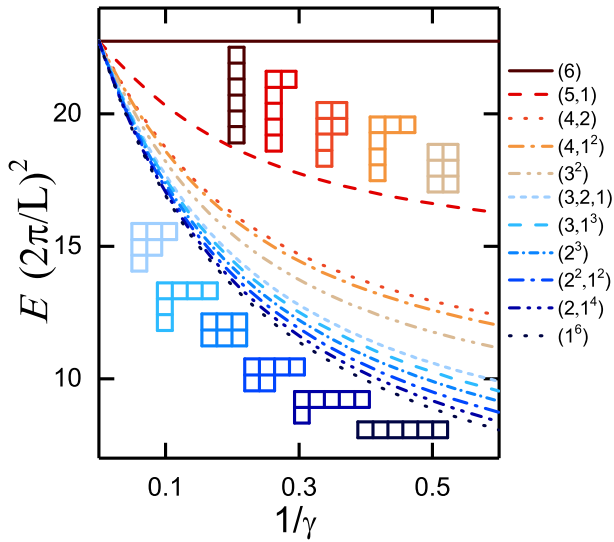}
\caption{(color online). The ground state energies of eleven symmetry
classes versus $1/\protect\gamma $ for the $SU(6)$ system with $N=6$. From
dark red solid line to dark blue dotted line correspond to the lowest energy
states of different symmetry classes described by the corresponding Young
tableaus.}
\label{fig4}
\end{figure}

%The conclusions are in agreement with the case of $SU(4)$ that there is no phase transition with the change of interaction strength and the order of ground state energy levels still holds on when the external potential changes from uniform to harmonic traps.

%\section{multi-component Fermi gases}
%\textit{Example of the }$SU(6)$\textit{\ system.---}

 Our results can be
directly generalized to the multi-component $SU(n)$ system with larger $n$,
for example, the system with $SU\left( 6\right) $ symmetry, for which the
BAEs take the form of Eqs. (\ref{log1}) and (\ref{log2}) with $n=6$. For the
example system with $N=6$, by solving the corresponding BAEs, we can get the
ground state energies for eleven symmetry classes. The corresponding results
are shown in Fig.\ref{fig4}. Compared to the case of $SU\left( 4\right) $,
there are two extra symmetry classes $(2, 1^4)$ and $(1^6)$, and similarly there also exist incomparable symmetry classes by
the pouring principle, e.g., $(4,1^{2})$ and $(3^{2})$, as well as $(3,1^{3})
$ and $(2^{3})$.
%Nevertheless, our results demonstrate that the order of ground state energy levels of different symmetry classes is also solely determined and can be characterized by their Young tableaux
The exact BA result gives the order of ground state energy levels of different symmetry classes: $E(6)\geq
E(5,1)\geq E(4,2)\geq \cdots \geq E(2,1^{4})\geq E(1^{6})$, where
\textquotedblleft $=$" holds true only in the TG limit $\gamma \rightarrow
\infty $. As shown in Fig.4,  the order is unchanged in the whole
interaction region.

\section{SUMMARY}
%\textit{Summary.---}
In summary, based on the BA solution of few-particle systems, we have
studied the ordering of energy levels for all kinds of permutation symmetry
classes of 1D multi-component Fermi systems with $SU(n)$ symmetry. In the
strongly interacting regime, from the expansion of the BA solutions, we
demonstrate that the system can be effectively described by an $SU(n)$ spin
exchange model with the exchange parameter being exactly determined.
Furthermore, the ordering of energy levels of the strongly interacting system trapped
in a harmonic potential is also determined by solving its effective spin-exchange model.

\begin{acknowledgments}
The work is supported by the National Key Research and Development Program
of China (2016YFA0300600), NSFC under Grants No. 11425419, No. 11374354 and
No. 11174360, and the Strategic Priority Research Program (B) of the Chinese
Academy of Sciences (No. XDB07020000). Y Z is supported by NSF of China
under Grant Nos. 11474189 and 11674201.
\end{acknowledgments}

\end{document}